# Racetrack microresonator based electro-optic phase shifters on a 3C-silicon-carbide-on-insulator platform

Tianren Fan[1], Xi Wu[1], Sai R. M. Vangapandu[1], Amir H. Hosseinnia[1], Ali A. Eftekhar[1], Ali Adibi[1,*]

[1]School of Electrical and Computer Engineering, Georgia Institute of Technology, Atlanta, GA 30332, USA

(Dated: February 10, 2021)

We report the first demonstration of integrated electro-optic (EO) phase shifters based on racetrack microresonators on a 3C-silicon-carbide-on-insulator (SiCOI) platform working at near-infrared (NIR) wavelengths. By applying DC voltage in the crystalline axis perpendicular to the waveguide plane, we have observed optical phase shifts from the racetrack microresonators whose loaded quality (Q) factors are ~ 30,000. We show voltage-length product ($V_\pi \cdot L_\pi$) of 118 V·cm, which corresponds to an EO coefficient $r_{41}$ of 2.6 pm/V. The SiCOI platform can be used to realize tunable SiC integrated photonic devices that are desirable for applications in nonlinear and quantum photonics over a wide bandwidth that covers visible and infrared wavelengths.

## I. INTRODUCTION

Silicon carbide (SiC) has recently attracted considerable attention in the fields of integrated photonics and quantum information science. Besides its wide bandgap, excellent quantum properties, and good optical nonlinearities, SiC is also a material proven to enable integrated photonic devices that can process optical signals on-chip [1-19]. Among all the integrated photonic devices, tunable optical devices including integrated optical modulators and phase shifters are essential elements as they convert electric signals to optical signals as well as trim the properties of the devices for some wavelength-sensitive applications.

Integrated optical modulators and phase shifters based on the free-carrier plasma dispersion effect have been developed using materials such as silicon (Si) [20], two-dimensional (2D) materials (e.g., graphene, transition metal dichalcogenides (TMDC), and black phosphorus) [21, 22], and several III-V semiconductor compounds, such as indium phosphide (InP) [23], and indium gallium arsenide phosphide (InGaAsP) [24]. The thermo-optic effect is also commonly used to realize phase shifters through the integration of metallic microheaters [11]. Pockels effect or electro-optic (EO) effect has many advantages over the above options for tunable integrated photonic devices, including faster tuning speed and lower power consumption. Operation of quantum photonic chips at the cryogenic temperatures rules out the use of plasma dispersion effect (as free carriers freeze) and the thermo-optic effect (due to extremely large power consumption), thus making the EO effect the best candidate for forming tunable and active photonic devices.

Several integrated active EO devices have been demonstrated with different materials during the past few years. Integrated lithium niobate (LN) EO modulators have been developed with data rates up to 210 Gbps and on-chip optical loss of < 0.5 dB [25]. Aluminum nitride EO modulators and phase shifters have also been demonstrated [26, 27]. However, CMOS-compatibility, fabrication challenges, and the possibility of integration with electronic functionalities on a single substrate are major shortcomings of these platforms.

Bulk 3C-SiC crystals demonstrate the EO effect with an EO coefficient $r_{41}$ of 2.7 pm/V [28] that can enable the modulation of the phase of an optical wave in a photonic waveguide by applying an electric field. Previously, the 3C-SiC EO modulators have been studied both numerically in the integrated form [29] and experimentally in the non-fully integrated form [30]. A major challenge in the experimental demonstration of fully-integrated SiC EO devices with practical performance measures has been the lack of a high-quality SiC integrated-photonic platform.

Our previous efforts in developing a low-cost high-quality 3C-silicon-carbide-on-insulator (SiCOI) platform and demonstrating optical microresonators with record-high quality (Q) factors [7, 15] have laid down the foundation for the material platform and the essential devices. Moreover, our previous demonstration of integrated thermo-optic phase shifters on SiCOI for effective tuning of optical phase in SiC microring resonators [11] confirms the possibility of vertical integration of electrical and photonic components in the SiCOI platform.

In this paper, we demonstrate the first integrated EO phase shifter on a SiCOI platform, to the best of our knowledge, by utilizing high-Q racetrack microresonators. The SiCOI sample is prepared through a low-temperature hydrophilic bonding process followed by chemical-mechanical polishing (CMP) as introduced in our previous work [7]. The SiCOI platform enables racetrack resonators with loaded Qs of ~ 30,000, which allows us to easily characterize the optical phase shifts by measuring the shifts in the resonance wavelengths under different electric fields (by applying different DC voltages). From the resonance shifts we extract a voltage-length product ($V_\pi \cdot L_\pi$) of 118 V·cm, which corresponds to an EO coefficient $r_{41}$ of 2.6 pm/V. It is also the first time the EO coefficient has been measured on the thin-film 3C-SiC, which is grown on the Si substrate originally via epitaxy.

## II. RELATED WORK

A typical design of the SiC EO phase-shifter section is shown in Fig. 1 (a). The propagation direction in the SiC waveguide is the [1 -1 0] crystalline direction of SiC with one contact (cathode) on the top and two contacts (anodes) with equal distances on the two sides of the waveguide. To avoid optical loss introduced by the metal contacts, there is a silicon oxide ($SiO_2$) cladding between the contacts and the waveguide. The architecture is a rib waveguide with a height of 650 nm, an etch depth of 450 nm, and a width of 800 nm to maintain a single mode. The cladding thickness is chosen as 900 nm to minimize the loss from contacts while allowing a reasonable value of the electric field in the vertical direction. The propagation loss of the waveguide simulated from gold contacts is about 0.004 dB/cm, which is negligible.

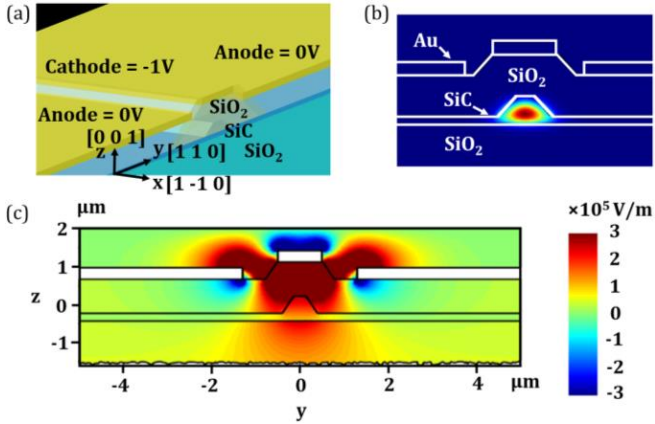

Fig. 1. (a) Schematic of the SiC EO phase shifter section. The SiC rib waveguide is along the [1 -1 0] direction with a SiO$_2$ cladding. There are three metal contacts on top of the cladding – one cathode on top of the waveguide and the other two anodes on the two sides of the waveguide, which can provide an electric field in the [0 0 1] direction by applying a voltage between the anodes and the cathode. (b) 2D TE$_0$ mode profile of the phase-shifter waveguide in (a). The waveguide has a width of 800 nm, a height of 650 nm, an etch depth of 450 nm, and a cladding thickness of 900 nm. The optical loss from gold (Au) contacts is 0.004 dB/cm simulated by Lumerical MODE FDE solver. (c) 2D amplitude profile of the electric field in the z ([0 0 1]) direction (E$_z$) of the phase shifter waveguide in (a) by applying -1 V between the cathode and the anodes. The distance between the cathode and each anode is 1 µm. This results in an average electric field of ~ 0.15 MV/m (computed by the Lumerical CHARGE Module) inside the waveguide in the area where the main extent of the TE$_0$ mode is.

By applying a voltage between the cathode and the two anodes in Fig. 1 (a), a vertical electric field (E) is generated in the [0 0 1] direction to enable the EO effect, governed by

$$\Delta n = \frac{n_{SiC}^3}{2} r_{41} E. \quad (1)$$

Here, $\Delta n$ is the change of the refractive index in [1 1 0] direction and $n_{eff}$ is the refractive index of SiC. Note that some stray field also exists in the [1 1 0] direction, with no change in the refractive index.

The best design strategy for the structure in Fig. 1 (a) is to put the anode right below the waveguide to form a capacitor-like structure, which maximizes the electric field in the vertical direction. However, the resulting fabrication process is not CMOS-compatible as it involves backside processing. On the other hand, the anodes can be placed directly next to the waveguide (through vias) and be as close to the cathode as possible in the horizontal direction to maximize the vertical electric field. In our experiments, for the consideration of the simplicity and the yield of the fabrication, we put all the contacts directly on the cladding.

Figure 1 (b) shows the two-dimensional (2D) field profile of the fundamental transverse-electric (TE$_0$) mode (i.e., electric field in the [1 1 0] (y) direction) together with the mentioned design parameters. The choice of TE$_0$ mode with its electric field lies mainly in the [1 1 0] direction enables the use of the refractive index change due to the EO effect while allowing for single-mode waveguide operation. Another design consideration is the overlap (Γ) between the applied electric field in the vertical direction and the optical mode in the horizontal direction [27]. Figure 1 (c) shows the profile of the applied electric field in the [0 0 1] direction in Fig. 1(a) when 1V is applied between the cathode and the two anodes resulting in an average electric field of ~ 0.15 MV/m in the region where the TE$_0$ mode has most of its energy and the field overlap Γ is ~ 0.2. Using Eq. (1) and the EO coefficient in Ref. [28], we can estimate a theoretical value of $V_\pi \cdot L_\pi$ of 113 V·cm. The relatively large value is caused by the relatively small EO coefficient of SiC compared to other EO materials like LN [25]. Nevertheless, smaller values of $V_\pi \cdot L_\pi$ might be achieved by further optimizing the device structure, specifically the contacts.

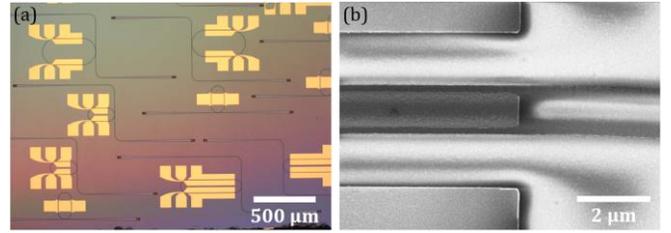

Fig. 2. (a) Optical micrograph of the fabricated SiC EO phase shifters. Racetrack microresonators with different roundtrip lengths are included to study the EO performance. The contact pads have a pitch of 100 µm, which matches that of the standard electrical probes. (b) SEM image of the electrodes on the phase-shifter section. The gaps between electrodes are cleared due to the successful liftoff process.

The fabrication process of the SiC EO phase shifter consists of two major steps – passive SiC device fabrication and contact fabrication. The passive SiC device fabrication process is similar to the one used in our previous work [11]. After electron beam lithography (EBL) and dry etching, there is a step of SiO$_2$ deposition via plasma-enhance chemical vapor deposition (PECVD) to clad the device. Since PECVD deposition is conformal, the cladding of the waveguide has a step height that is almost the same as the height of the rib waveguide, which is 450 nm. After the cladding deposition, the contacts are fabricated onto the top of the cladding using EBL with PMMA A6 as the mask, sputtering of Au and titanium (Ti), and the metal liftoff in the organic solvent acetone. The thickness of PMMA A6 is chosen as 700 nm to be higher than the step height of the waveguide profile (i.e., 450 nm). Since some parts of the contacts go through this step height, we use sputtering to deposit metals as it can cover all the profile of the waveguide with its partial conformality (compared with other deposition techniques such as electron beam evaporation). To achieve sub-µm feature sizes for the top contacts and the gap between the contacts, we deposit 10 nm Ti followed by 100 nm Au (note that the mask depth could be as low as 250 nm on top of the waveguide). The liftoff process consists of an overnight soak in the acetone followed by 5 minutes of ultrasonication to fully remove the metal in the unwanted area, especially in the gaps between contacts. Figure 2 (a) shows the optical micrograph of the fabricated SiC EO devices, including racetrack microresonators with different roundtrip lengths, to study the EO performance. The fabrication of different components is successful with defined clear boundaries and fewer particles or damages. Figure 2 (b) shows the scanning electron microscopy (SEM) image of the electrodes on the phase-shifter section.

The racetrack resonators in Fig. 2 (a) are designed with the straight part of the racetrack lying along with the contacts in the [1 1 0] or [1 -1 0] direction. The optical mode that propagates along the straight part of the resonator experiences an optical phase shift due to the EO effect. On the other hand, the microresonator provides a narrow linewidth in the spectral domain due to its high-Q feature. This facilitates the observation of the phase change without a full $2\pi$ phase shift.

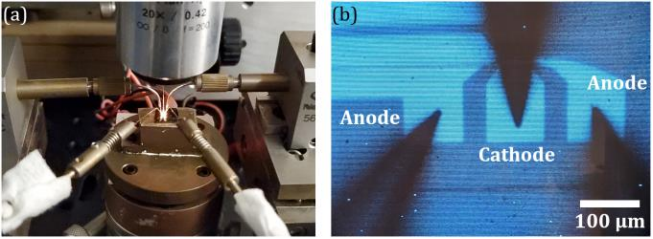

Fig. 3. (a) Photo of the characterization setup used to measure the EO phase-shifter devices. Device characterization has two major parts: 1) the optical measurement of the passive photonic devices through cleaved fibers, and 2) the electrical measurement of the tunable photonic devices through electrical probes. (b) Micrograph from the top camera showing a SiC EO phase shifter device under probing conditions.

The EO devices are characterized using a setup shown in Fig. 3 (a), which enables both passive (optical) device measurements and electrical tuning measurements. The optical measurements are performed by coupling light from a tunable laser (Agilent Technologies 81680A, linewidth 100 kHz) operating in the continuous sweep mode to the input grating on the device through a cleaved optical fiber roughly aligned at 8° and collecting the output signal from the output grating through another cleaved fiber and sending it to a photodetector (Thorlabs PDB150C). The electrical setup contains three DC probe stages and a source measurement unit (SMU, Keithley 2614B). Figure 3 (b) shows the view of the probing experiment through the microscope: one probe is in contact with the cathode while the other two probes are in contact with the anodes. The SMU applies a voltage between the cathode and each anode while monitoring the leakage current to avoid the device breakdown. During each measurement, the resonance position is calibrated using a passive resonant device.

Figure 4 shows the EO measurement results for two racetrack resonators with different roundtrip lengths. The first device (see Fig. 4 (a)) has a roundtrip length of 622 μm with an EO phase shifter section of 123 μm. The FSR of the resonances is measured to be 1.42 nm. The distance between the cathode and each anode is 1 μm. The transmission spectra under different voltages for the structure in Fig. 4 (a) are shown in Fig. 4 (b). From the transmission spectra (10.6 pm resonance shift over 160 V), we can extract $V_\pi \cdot L_\pi$ of 132 V·cm. The second device (see Fig. 4 (c)) has a roundtrip length of 622 μm with an EO phase shifter section of 217 μm and FSR of 1.42 nm. The corresponding transmission spectra under different voltages (31.3 pm resonance shift over 240 V) are shown in Fig. 4 (d) demonstrating $V_\pi \cdot L_\pi$ of 118 V·cm. The $V_\pi \cdot L_\pi$ of the EO devices are close to each other and match with the theoretical prediction of 113 V·cm. We think this disparity between the experimental and theoretical results might be caused by two sources: 1) the fabrication error (e.g., the curved (instead of flat) shape of the top contact, the variation of the feature sizes due to the lithography variations), and 2) the lower EO coefficient of the SiC film used in the experiments compared to the bulk value (3C-SiC tends to be polycrystalline, which leads to a smaller EO coefficient). The actual EO coefficient of the thin film SiC used in the experiment is extracted to be 2.6 pm/V. Our value is close to the one measured from a bulk 3C-SiC crystal [28], which shows our CMP process effectively produces high-quality 3C-SiC thin films [7,15].

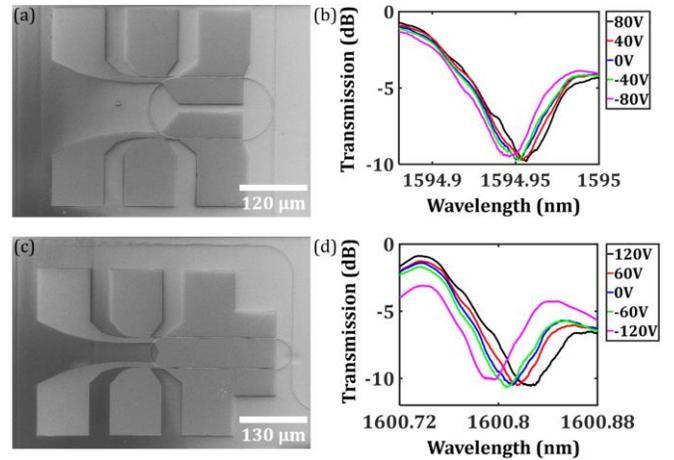

Fig. 4. (a) SEM image of a SiC EO phase shifter based on a racetrack microresonator with 60 μm-radius waveguide bends and 123 μm straight (phase-shifter) waveguide sections. (b) Transmission spectra of the racetrack microresonator in (a) with different voltages applied between the cathode and the anodes. The linear response of the resonance shift versus voltage is observed. (c) SEM image of a SiC EO phase shifter with a racetrack microresonator with 30 μm-radius bends and 217 μm straight sections. (d) Transmission spectra of the racetrack microresonator in (c) with different voltages applied between the cathode and the anodes. Again, the linear response of resonance shift versus voltage is observed.

The performance of the EO phase shifter can be further improved by improving the material quality and optimizing the device structure and fabrication process. A straightforward approach is to fabricate the contacts for the anodes before the cladding and make vias through the cladding on the two sides of the waveguide to maximize the overlap between the vertical electric field and the optical mode. On the other hand, to improve the Q-factors of our microresonators, we can further improve the material quality by using a high-quality thick (~ 5 μm) SiC film to make the SiCOI platform as discussed in our previous work [15]. Finally, an aggressive device optimization can further improve $V_\pi \cdot L_\pi$.

## III. CONCLUSION

In conclusion, we demonstrated here the first SiC integrated optical phase shifters using the SiCOI platform that we introduced to the field, for the first time [7]. Our experiments show tuning capability $V_\pi \cdot L_\pi$ of 118 V·cm. It is also the first time that the EO coefficient $r_{41}$ of a thin-film epitaxial 3C-SiC has been measured, with a value of 2.6 pm/V. We believe that we could further improve the performance of the optical phase shifter by improving the material quality and optimizing the device structure and fabrication process. This work would pave the way for integrated photonics applications on 3C-SiC in both classical and nonclassical areas over a wide bandwidth.

**Funding:** This work was supported by the Air Force Office of Scientific Research (AFOSR) Grant no. FA9550-15-1-0342 and the Office of Naval Research Grant no. N00014-15-1-2081. This work was performed in part at the Georgia Tech Institute for Electronics and Nanotechnology (IEN), a member of the National Nanotechnology Coordinated Infrastructure (NNCI), which is supported by the National Science Foundation (ECCS-1542174).